\documentclass[pra,reprint,doublecolumn,superscriptaddress, nobalancelastpage]{revtex4-1}

\usepackage{tikz}

\usepackage{etoolbox}
\patchcmd{\section}
  {\centering}
  {\raggedright}
  {}
  {}
\patchcmd{\subsection}
  {\centering}
  {\raggedright}
  {}
  {}

\usepackage{tcolorbox}
\usepackage[sort&compress]{natbib}
\usepackage{soul}
\usepackage{physics}
\usepackage{graphicx}
\usepackage{epstopdf}
\usepackage{verbatim}
\usepackage{float}
\usepackage{sidecap}
\sidecaptionvpos{figure}{c}
\usepackage{lipsum}
\usepackage{capt-of}
\usepackage{setspace}
\usepackage{dcolumn}
\usepackage{bm}
\usepackage{amssymb}
\usepackage{amsmath}
\usepackage{mathtools}
\usepackage{color}
\usepackage{multirow}
\usepackage{lipsum}
\usepackage[a4paper]{geometry}
\newgeometry{top=1.25cm,right=1cm,left=1.5cm,bottom=1.25cm}
\usepackage[colorlinks=true, urlcolor=blue, pdfborder={0 0 0}]{hyperref}
\hypersetup{
linkcolor=blue,
citecolor=blue
}

\draft

\begin{document}

\title{\large{\textbf{Influence of sample momentum space features on scanning tunnelling microscope measurements} }}

\author{Maxwell T. West} \affiliation{Center for Quantum Computation and Communication Technology, School of Physics, University of Melbourne, Parkville, 3010, VIC, Australia.}


\author{Muhammad Usman} \email{musman@unimelb.edu.au} \affiliation{Center for Quantum Computation and Communication Technology, School of Physics, University of Melbourne, Parkville, 3010, VIC, Australia.} \affiliation{School of Computing and Information Systems, Melbourne School of Engineering, University of Melbourne, Parkville, 3010, VIC, Australia}
\maketitle

\onecolumngrid

\noindent
\textcolor{black}{
\normalsize{\textbf{Theoretical understanding of scanning tunnelling microscope (STM) measurements involve electronic structure details of the STM tip and the sample being measured. Conventionally, the focus has been on the accuracy of the electronic state simulations of the sample, whereas the STM tip electronic state is typically approximated as a simple spherically symmetric $ s $ orbital. This widely used $ s $ orbital approximation has failed in recent STM studies where the measured STM images of subsurface impurity wave functions in silicon required a detailed description of the STM tip electronic state. In this work, we show that the failure of the $ s $ orbital approximation is due to the indirect band-gap of the sample material silicon (Si), which gives rise to complex valley interferences in the momentum space of impurity wave functions. Based on direct comparison of STM images computed from multi-million-atom electronic structure calculations of impurity wave functions in direct (GaAs) and indirect (Si) band-gap materials, our results establish that whilst the selection of STM tip orbital only plays a minor qualitative role for the direct band gap GaAs material, the STM measurements are dramatically modified by the momentum space features of the indirect band gap Si material, thereby requiring a quantitative representation of the STM tip orbital configuration. Our work provides new insights to understand future STM studies of semiconductor materials based on their momentum space features, which will be important for the design and implementation of emerging technologies in the areas of quantum computing, photonics, spintronics and valleytronics.}}}
\\ \\ \\
\twocolumngrid

\noindent
An important aspect of modern materials science and engineering is the ability to place impurities into semiconductors with nanometre precision~\cite{Koenraad_NMaterials_2011, Fuechsle_NN_2012, Weber_Science_2012, Ho_NM_2008, SAE_2013}. These impurities drastically modify the band structure properties of their host materials, leading to novel electronic, optoelectronic and quantum properties suitable for a diverse range of nanoscale devices working in both classical \cite{Pierre_NNano_2010, Sarkar_Nature_2015, Ionescu_Nature_2011} and quantum \cite{Kane_Nature_1998, Hill_science_2015} regimes of operation. The design and engineering of impurity atoms in semiconductor materials, however, demand high precision fabrication and characterisation, often with atomic resolution, which is a challenging task. Scanning tunnelling microscope (STM) has been one of the most useful and widely used tools, which offers unprecedented capabilities to manipulate and characterise nanomaterials down to single atom resolution \cite{Fuechsle_NN_2012, Weber_Science_2012}. Since its invention in 1981 at IBM and the subsequent physics Nobel prize in 1986, STM has been extensively used to design a wide range of materials including semiconductors \cite{Koenraad_NMaterials_2011, Fuechsle_NN_2012, Weber_Science_2012, Garleff_PRB_2008, Marczinowski_PRB_2008, gaas_stm_2015, gaas_stm2_2017}, 2D materials \cite{Jiang_Research_2019}, organic molecules \cite{Gross_PRL_2011}, and metal-organics \cite{Marina_Small_2021}. Recently, STM has been used to probe the electronic structure properties of individual impurity atoms in semiconductors by producing high resolution spatially-resolved images of bounded single \cite{Salfi_NatMat_2014, Usman_NN_2016} and coupled \cite{Voisin_2020} electron wave functions. The theoretical modelling of these STM images not only offers pathways to gain an exquisite understanding of the fundamental impurity physics \cite{Usman_Nanoscale_2017}, but also leads to the design of precision metrology techniques \cite{Usman_NN_2016}, capability to characterise qubits at large-scale~\cite{Usman_NPJCM_2020}, and mapping of the interactions between the coupled electron states \cite{Voisin_2020}. Therefore, an accurate theoretical understanding of STM measurements is important to fully exploit the capabilities of this highly versatile tool for the advancement of nanomaterial science and engineering, enabling new technologies with a wide range of applications in the areas of photonics and quantum computing. 

\begin{figure*}[htbp]
\begin{center}
\includegraphics[width=19cm]{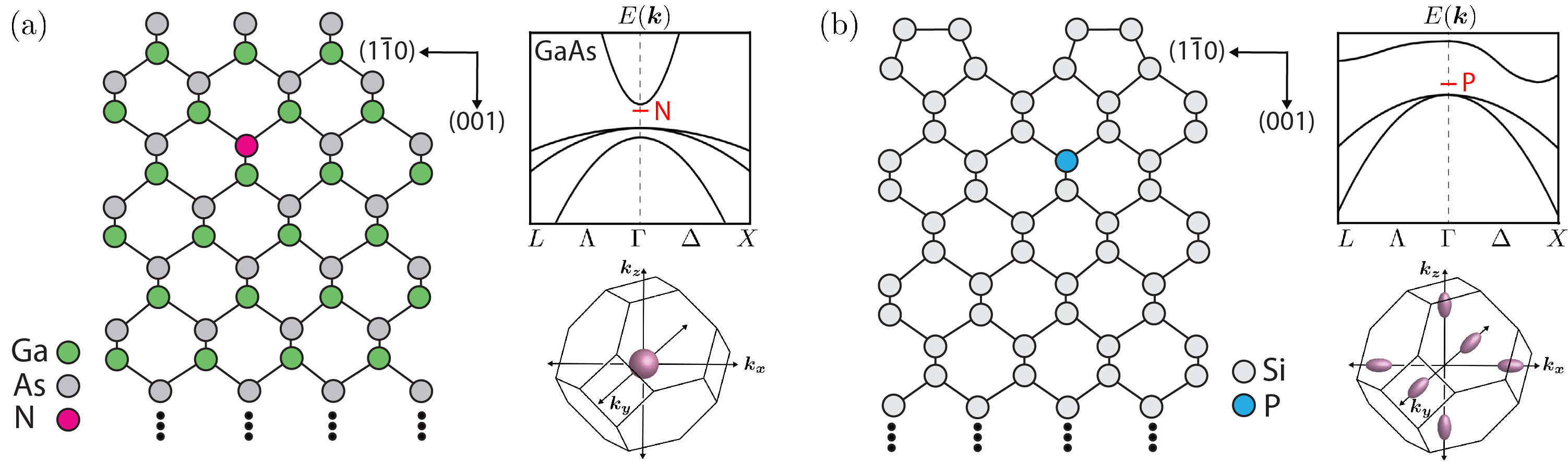}
\caption{Schematic representation of a nitrogen (N) impurity atom in the GaAs lattice is shown in (a), and a phosphorus (P) impurity atom in the Si lattice is shown in (b). Illustrations of the respective GaAs and Si band structures are also shown along with the $L-\Gamma-X$ path through the reciprocal space. GaAs has a direct band gap, and Si has an indirect band gap with conduction band minima along the six equivalent (100) directions, about 82\% of the way to the Brillouin zone boundary. The impurity states for both GaAs and Si sit just below the conduction band as indicated by red color markers. In momentum space, the wave function associated with a N state in GaAs is concentrated around the $\Gamma$ point, but due to the indirect Si band structure, the P donor state resides in a superposition of the six-fold degenerate valleys along the six energy minimising directions. }
\label{fig:1}
\end{center}
\end{figure*} 

The theoretical understanding of STM measurements involves the calculation of electronic structure of STM tip (usually made up from a transition metal such as tungsten) and sample (the system being investigated). Much of the focus has been on accurate simulations of the sample electronic states, and relatively little attention has been given to the role of STM tip electronic state. As the determination of the exact electronic structure of an STM tip is a challenging task, it is often approximated by a single $s$ orbital based on the formalism developed by Tersoff and Hamann \cite{Tersoff_PRB_1985}. In the Tersoff-Hamann model, the tunnelling current from a sample taken at a bias voltage $V$ is given by the integral through an energy window $eV$ of the local density of states (LDOS) of the impurity evaluated at the tip apex, which can be simplified as sample charge density vacuum decayed at the tip apex. This approximation has been successful for many STM based investigations, providing a very good qualitative understanding of the measured STM datasets, albeit in contradiction with an earlier theoretical prediction, which stated that the STM measurements should be dominated by $d_{z^2-r^2/3}$ type orbital in the case of a transition metal STM tips \cite{Chen_PRB_1990}. \\ 

A recent study of phosphorus impurities in silicon (Si:P) exhibited a very strong dependence of STM measurements on the tip electronic structure \cite{Usman_NN_2016}. The P impurities were placed at various lattice positions up to 5 nm below the silicon surface and in each case, STM images of single electron wave functions bounded to impurity atoms showed drastically different symmetry and brightness of features computed based on tip orbital selection. Remarkably, it was also shown that the computed STM images were in excellent agreement with the measured images only when the tip orbital consisted of a dominant $d_{z^2-r^2/3}$ orbital, consistent with the earlier prediction for a transition metal tip \cite{Chen_PRB_1990}. This behaviour was in stark contrast to the observed STM images of impurities in direct band-gap semiconductor materials such as GaAs \cite{gaas_stm_2015, gaas_stm2_2017} and InP \cite{inp_stm_2017}, where the images of electrons bounded to subsurface Bismuth (Bi), Nitrogen (N), and Antimony (Sb) impurities at various depths were in good agreement with theory based on only $ s $ orbital in the tip state. 

In this work, we show that the failure of the Tersoff-Hamann $ s $ orbital approximation for the case of Si:P is a result of the indirect band structure of silicon, and its associated six-fold valley degeneracy, which introduces complex momentum space interferences leading to rich high frequency components. This is established by directly comparing STM images from Si:P indirect band gap system with the GaAs:N direct band gap system, where the calculations of STM images are based on multi-million-atom tight-binding simulations of impurity wave functions \cite{Usman_JPCM_2015, Usman_PRAppl_2018}, coupled with the Bardeen's tunnelling theory \cite{Bardeen_PRL_1961} and Chen's derivative rule \cite{Chen_PRB_1990}. The underlying physics is further investigated by means of a simple Kohn-Luttinger model for impurity wave function \cite{Kohn_PR_1955}. The KL model produces a less accurate approximation, only exhibiting qualitative features of the experimentally observed STM images \cite{Usman_NN_2016}, but has a simple analytic form which is more amenable to theoretical analysis than the purely numerical output of the detailed tight-binding simulation. In particular, we can artificially tune the values of the valley wave vectors, interpolating between the case of an indirect and a direct band gap material. \\

\begin{figure*}[htbp]
\begin{center}
\includegraphics[width=13cm]{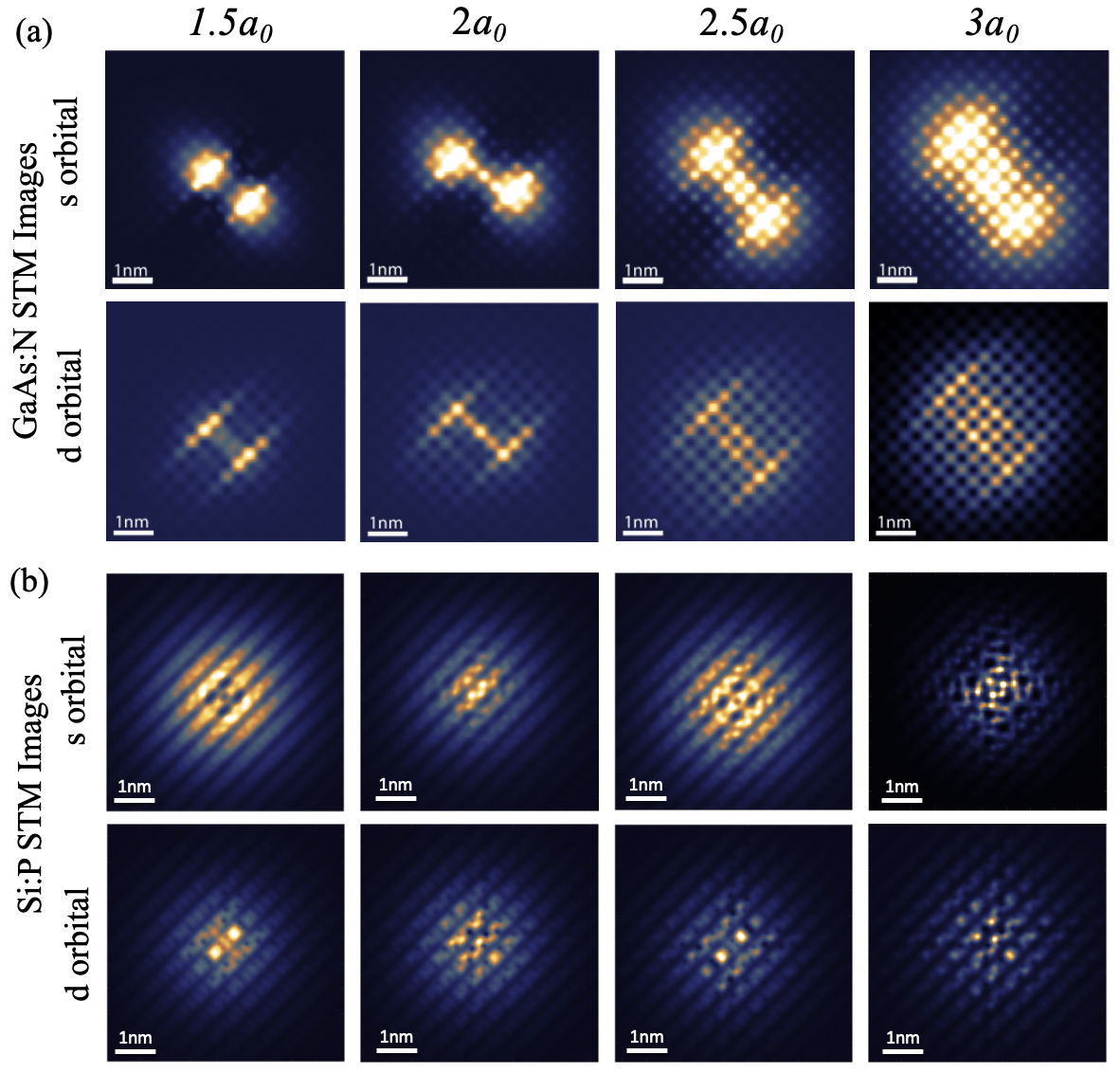}
\caption{The computed STM images of impurity wave functions are shown at various depths as calculated by multi-million atom nearest-neighbour tight binding simulations, coupled with the Bardeen's tunnelling formalism and Chen's derivative rule. The $s$ and $d_{z^2-r^2/3}$ (abbreviated just as $d$) orbital images are calculated by using Chen's derivative rule. All images are plotted by using the same color scale. $ a_{ 0}  $ is the lattice constant of the host semiconductor material. }
\label{fig:2}
\end{center}
\end{figure*}

For the purpose of this study, we have investigated two material systems: Nitrogen (N) impurities in a direct band-gap material GaAs and Phosphorus (P) impurities in an indirect band-gap material silicon (Si). The set-up is schematically shown in Figure \ref{fig:1}. Si and GaAs share the same bulk crystal lattice structure, with the primary structural difference between the two cases being the 2$\times$1 surface reconstruction of Si at the vacuum interface, which is typically the case in experimentally fabricated samples \cite{Fuechsle_NN_2012}. The relevant details of their respective band structures are also depicted in Figure \ref{fig:1}. The GaAs material being a direct band-gap material has minimum (maximum) of the conduction (valence) band at the $\Gamma$ point. The N impurity related electronic energy level is directly under the conduction band minimum as shown by a red marker. Contrarily, Si material being an indirect band-gap material has valence band maximum at the $\Gamma$ point, but the conduction band minimum is around 82\% of the way to the Brillouin zone boundary along the $X$ direction. The phosphorus related impurity energy level is below the Si conduction band minimum inside the band-gap region \cite{Usman_JPCM_2015}. In the Fourier space, the N impurity state in GaAs is therefore localised on the $\Gamma$ point (Figure \ref{fig:1} (a)), while the P donor impurity state in Si sits in a superposition of the six valleys, corresponding to the six conduction band minima along the $X$ direction (Figure \ref{fig:1} (b)). For a P atom placed in the bulk Si, the ground impurity state is a spin degenerate singlet with equal contributions from all six valleys \cite{Usman_PRB_2015}. However, when the donor is located closer to the silicon-vacuum interface, the valleys no longer contribute equally to the ground state; instead, there is re-population from the  $  \pm x $ and  $  \pm y $ valleys into the $  \pm z $ valleys, which will be discussed later in this work.

To investigate STM images of impurity wave functions in GaAs and Si materials, we begin by simulating the electronic structure of GaAs:N and Si:P systems where a single impurity atom (P or N) is placed in the host material (Si or GaAs) at selective sites below the vacuum interface. In each case, the electronic structure calculations are based on multi-million-atom tight-binding simulations, and the details of the established tight-binding models for GaAs and Si materials are reported in our earlier studies \cite{Usman_JPCM_2015, Usman_PRB_2015, Usman_PRAppl_2018}. The tight-binding models are benchmarked against the available experimental datasets to accurately reproduce the energy levels of P and N impurities in their respective Si and GaAs host materials under bulk conditions. The simulation domain in each case consists of about 4 million atoms. For Si:P system, the $z$=0 surface is 2$\times$1 reconstructed \cite{Craig_SS_1990, Usman_NN_2016}. The computation of the STM images is performed by coupling the atomistic tight-binding wave function calculation with the Bardeen's tunnelling formalism~\cite{Bardeen_PRL_1961}. The impurity wave functions are decayed in the vacuum region based on the Slater orbital real-space representation~\cite{Slater_PR_1954}. The effect of the STM tip state is studied by Chen's derivative rule \cite{Chen_PRB_1990}. In this study, we mainly focus on two STM tip orbitals, the widely used $ s $ orbital approximation and the $d_{z^2-r^2/3}$ orbital which was recently found important to match the experimental measurements of Si:P system. Moreover, the $d_{z^2-r^2/3}$ orbital is also relevant for transition metal STM tips which are most commonly used for nanomaterial studies. The $ s $ and $d_{z^2-r^2/3}$ orbital type images reflect different applications of the Chen's rule; in the $s$ case, the images are proportional to the square of the impurity state wave function evaluated at the tip apex \textit{i.e.} ${\textrm I_\textrm {T}} (r_0) = \lvert \Psi_{\rm D} \rvert _{r_0} ^2$, whereas in the $d_{z^2-r^2/3}$ case, the images are proportional to a linear combination of second derivatives of the impurity wave function, again evaluated at the tip apex: 

\begin{equation*}\label{func}
{\textrm I_\textrm {T}} (r_0) \varpropto \left\lvert \frac{2}{3}\frac{\partial^2 \Psi_ \textrm D (r)}{\partial z^2} - \frac{1}{3}\frac{\partial^2 \Psi_ \textrm D (r)}{\partial y^2} - \frac{1}{3}\frac{\partial^2 \Psi_ \textrm D (r)}{\partial x^{2} } \right\rvert _{r_0} ^2
\end{equation*}

\noindent where $\Psi_{\rm D}$ is the donor wave function and $r_0$ is the position of the STM tip.  \\

Figure \ref{fig:2} plots the computed STM images for GaAs:N and Si:P systems for a few impurity atom locations. In the case of GaAs:N (Figure 2 (a)), our results show considerable similarities between the $ s $ and $d_{z^2-r^2/3}$ orbital images. The images exhibit qualitatively similar symmetry and brightness of features, with $s$ orbital images being relatively blurred version of the $d_{z^2-r^2/3}$ case. However, in the presence of blurring noise typically present in experimental measurements \cite{Usman_NPJCM_2020}, this difference is expected to be slim. Notably, when we extract feature boundaries from the $d_{z^2-r^2/3}$ images and overlay them on the corresponding $s$ orbital images, the symmetry and size of the features is found to be in very good agreement (see supplementary Figure \ref{fig:S1}). Therefore, we conclude that for the GaAs:N system, the precise tip orbital composition does not play an important role, and the $s$ orbital Tersoff-Hamann approximation provides a qualitatively accurate understanding of the measurements. This has indeed been true in several recent studies where the computed $s$ orbital images were quite accurate to understand direct band-gap STM experiments \cite{gaas_stm_2015, gaas_stm2_2017, inp_stm_2017}. \\

Contrarily, the STM images corresponding to the Si:P system (Figure \ref{fig:2}(b)) show highly distinct features based on the tip orbital selection. The images computed at several depths show that neither symmetry nor the sizes of features match for $s$ and $d_{z^2-r^2/3}$ orbital configurations. Moreover, the images display complicated structures which are a strong function of the exact lattice position of P atom in Si \cite{Usman_NN_2016}. We attribute this stark difference between the $s$ and $d_{z^2-r^2/3}$ orbital images to the presence of momentum space valleys which lead to high frequency interference patterns. Indeed the supplementary Figure \ref{fig:S2} plots the Fourier space images for both GaAs:N and Si:P cases, indicating that the Fourier spectra for an Si:P image shows highly rich spectra. \\

\begin{figure}[htbp]
\begin{center}
\includegraphics[width=9.25cm]{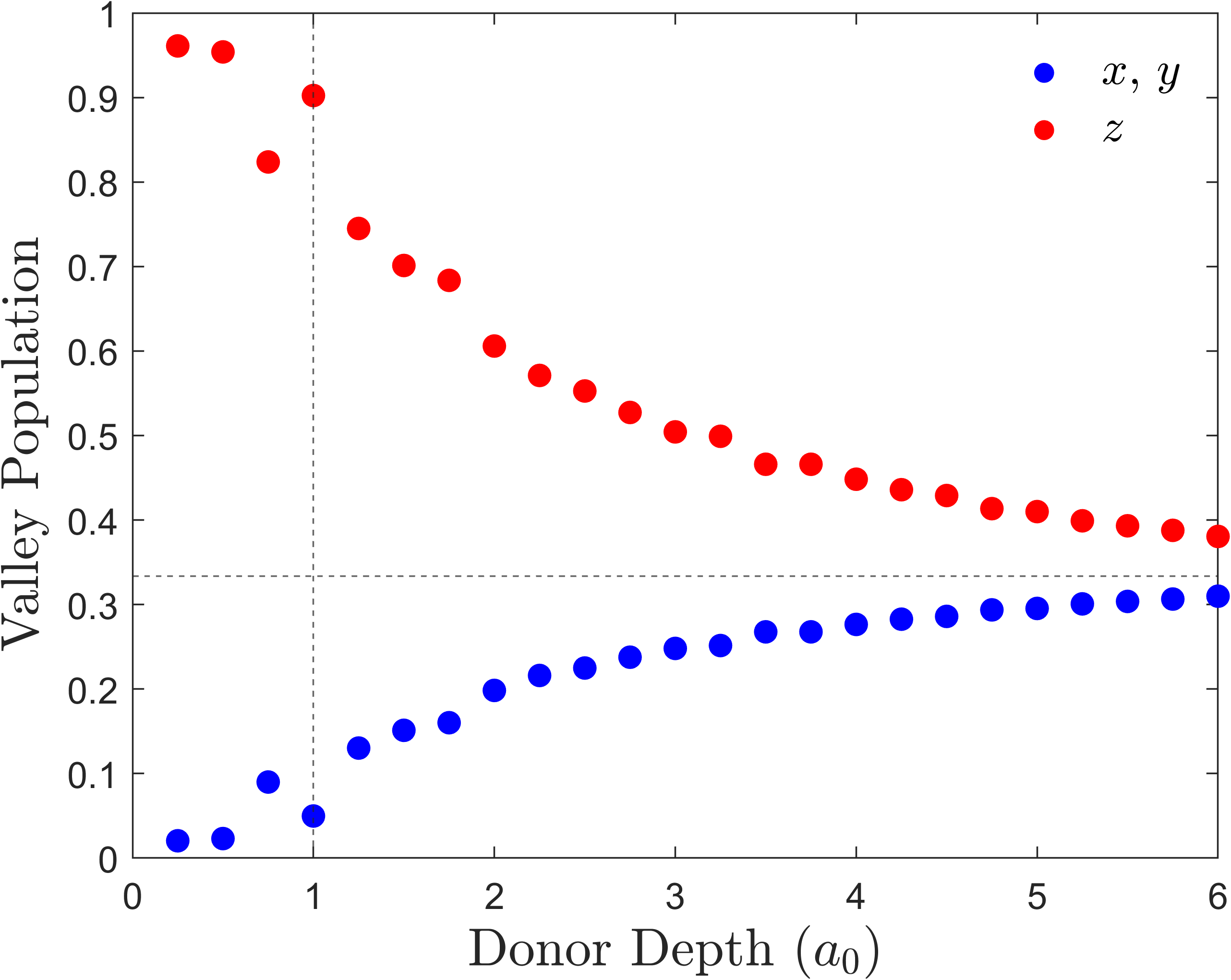}
\captionof{figure}{The valley population of phosphorus donor states in Si is plotted as a function of the depth from the (001) surface. The donor atoms closer to the surface experience significant interface and strain effects, leading to valley re-population from the $x$ and $y$ valleys to the $z$ valleys, which depends on both the depth of the donor and its lateral position with respect to the Si surface dimer rows. As the donor depth increases, the population of all valleys converges towards the bulk value of 1/3, plotted as a dotted horizontal line. The vertical dotted line at 1$a_0$ donor depth indicates that the significant valley repopulation below this depth will transform Si:P donor wave functions to nearly single $z$ valley states, leading to STM images similar to the single valley GaAs:N case. }
\label{fig:3}
\end{center}
\end{figure}

In order to further understand the role of valley configurations, we compute valley population of Si:P wave functions as a function of P atom depth from the vacuum interface, which is shown in Figure \ref{fig:3}. In the bulk case (large depths), a P donor in Si sits in an equal superposition of all six valleys (33\% population), and therefore the effect of valley interference is expected to be strong. Indeed, the $s$ and $d_{z^2-r^2/3}$ orbital STM images of deeper P donor depths exhibit stronger mismatch in Figure \ref{fig:2}(b). The supplementary Figure \ref{fig:S3} shows Si:P images for impurity depths approaching bulk limit, indicating that the different between $s$ and $d_{z^2-r^2/3}$ orbital images is more pronounced when depth is increased. However, when the P donor is closer to the vacuum interface, the effect of interface and reconstruction related strain leads to strong population of $z$ valleys at the expense of $x$ and $y$ valleys. For donor depths below $ a_{ 0} $, the $z$ valley population is more than 80\%. This leads to a weak difference between $s$ and $d_{z^2-r^2/3}$ orbital Si:P images as shown in supplementary Figure \ref{fig:S4}. This implies that the lattice incommensurate valley oscillations in the $ x $ and  $ y $ directions, which contribute to the rich structure of the STM images of deep donors, contribute significantly less to the wave functions of shallow donors, supporting our understanding that indeed the presence of valley related interferences enhance the role of tip electronic state in the calculation of STM images. This is also in agreement with a recent report on AlAs:Si system \cite{Tjeertes_arXiv_2021}, where the valley impact was found to be weak for impurities closer to the surface. \\

Another important feature of valleys is that STM images remain distinct for deep donor depths. Even at 5 nm (10 $a_0$) depth, the $s$ and $d_{z^2-r^2/3}$ orbital images retain symmetry and can be distinctly identified for each position of impurity atom. This was exploited in a recent study to develop an exact atom spatial metrology technique to pinpoint phosphorus donor atoms in silicon \cite{Usman_NN_2016}. However, in the case of the GaAs:N system, the wave functions STM images lack any distinct character of features and therefore will not allow spatial metrology at such deep depths. To illustrate this effect, the supplementary material Figure \ref{fig:S5} show the computed STM images for GaAs:N and Si:P systems when N and P atoms are placed at relatively deeper depths. \\

To gain further insight into the influence of the valley degeneracy on the STM image features, we consider a simple analytical model of the Si:P system by Kohn and Luttinger\cite{Kohn_PR_1955}, in which the donor ground state wave function takes the form:

\begin{equation*}
    \Psi_D(\boldsymbol{r})= \frac{1}{\sqrt{6}} \sum_{\mu} e^{i\boldsymbol{k}_\mu\vdot\boldsymbol{r}} F_{\mu} (\boldsymbol{r})  u_{\mu} (\boldsymbol{r})
\end{equation*}

\noindent
where the $F_{\mu}$ are envelope functions, the $u_{\mu}$ are the periodic Bloch functions and the $\boldsymbol{k}_{\mu}$ are the valley wave vectors, $\boldsymbol{k}_{\mu}\in 2\pi  (0.82)/a_0\{ \boldsymbol{k}_{x},\boldsymbol{k}_{y},\boldsymbol{k}_{z} \} $. Following the Kohn and Luttinger formalism, we take the ground state envelope functions to be elongated Gaussian, with width and length given by a pair $a,b$ of variationally determined effective Bohr radii, distinct due to the effective mass anisotropy of silicon. The periodic functions $u_\mu$ can be Fourier expanded as $u_{\mu}(\boldsymbol{r}) = \sum_{\boldsymbol{G}} A_{\mu,\boldsymbol{G}} e^{i \boldsymbol{r}\vdot\boldsymbol{G} }$, where the sum is over the reciprocal lattice vectors $\boldsymbol{G}$ and the $A_{\mu,\boldsymbol{G}}$ can be determined by a DFT method \cite{dft_coefficients_2011}. The majority of this decomposition of the $u_\mu$ is into a small number of terms, meaning that we can get an accurate approximation to the KL wave function by keeping only the first few $A_{\mu,\boldsymbol{G}}$, thereby obtaining a simple, analytic ansatz of the donor wave function. \\

The envelope functions $F_\mu$ of the KL donor wave function are taken to be of the form:

\begin{equation*}
    F_{\pm z}(\boldsymbol{r}) = \frac{1}{\sqrt{\pi a^2 b}}\exp\left( -\sqrt{ \frac{x^2+y^2}{a^2} + \frac{z^2}{b^2} } \right)
\end{equation*}

\noindent
and similarly for the $\pm x$ and $ \pm y$ valleys. The variational parameters $a$ and $b$ are taken from \cite{Saraiva_PRB_2015} to be 0.9 nm and 0.52 nm, respectively. The coefficients in the Fourier expansion $A_{\mu,\boldsymbol{G}}$ of the Bloch functions $u_\mu$ are taken from \cite{dft_coefficients_2011}, where they were calculated by first-principles density functional theory. Dropping the terms with $ \abs{A_{\mu,\boldsymbol{G}}}^2 \leq 2\cp 10^{-3} $ gives a closed form expression for the donor wave function:

\begin{align*}
\Psi_D(\boldsymbol{r})  &=  \exp \left( -\sqrt{ \frac{y^2+z^2}{a^2} + \frac{x^2}{b^2} } \right)  \Bigg[ 2A \cos\left( \frac{2\pi vx}{a_0}\right)\\
&-8B    \bigg( \cos\left( \frac{2\pi (1-v)x}{a_0}\right)\cos\left(\frac{2\pi y}{a_0}\right)\cos\left(\frac{2\pi z}{a_0}\right)\\
&+\sin\left(\frac{2\pi(1-v) x}{a_0}\right)\sin\left(\frac{2\pi y}{a_0}\right)\sin\left(\frac{2\pi z}{a_0}\right) \bigg)\\
&-4C \cos \left( \frac{2\pi (2-v) x}{a_0}\right) \bigg( \cos \left( \frac{4\pi y}{a_0}\right)+\cos \left(\frac{4\pi z}{a_0}\right) \bigg)  \\
&+8D    \bigg( \cos\left( \frac{2\pi (1+v)x}{a_0}\right)\cos\left(\frac{2\pi y}{a_0}\right)\cos\left(\frac{2\pi z}{a_0}\right)\\
&+\sin\left(\frac{2\pi(1+v) x}{a_0}\right)\sin\left(\frac{2\pi y}{a_0}\right)\sin\left(\frac{2\pi z}{a_0}\right) \bigg)   \Bigg]\\
&+ \mathrm{cyclic\ permutations\ of\ } x,y,z.
\end{align*}

\noindent
where we have introduced the ``valley parameter'' $v$ whose value varies from 0 to 1, where 0(1) corresponds to $\Gamma$($X$) points on band structure plot, and $v$ = 0.82 indicates the position of the lowest energy point in silicon conduction band. The valley parameter can be tuned to artificially change valley contributions in donor state $\Psi_D$,  and can be set to zero to simulate the effect of removing the valleys. The values of the coefficients are given by $A=0.3428,\ B = 0.3131,\ C = 0.0986$ and $D = 0.0695$. \\

\begin{figure*}[htbp]
\begin{center}
\includegraphics[width=12cm]{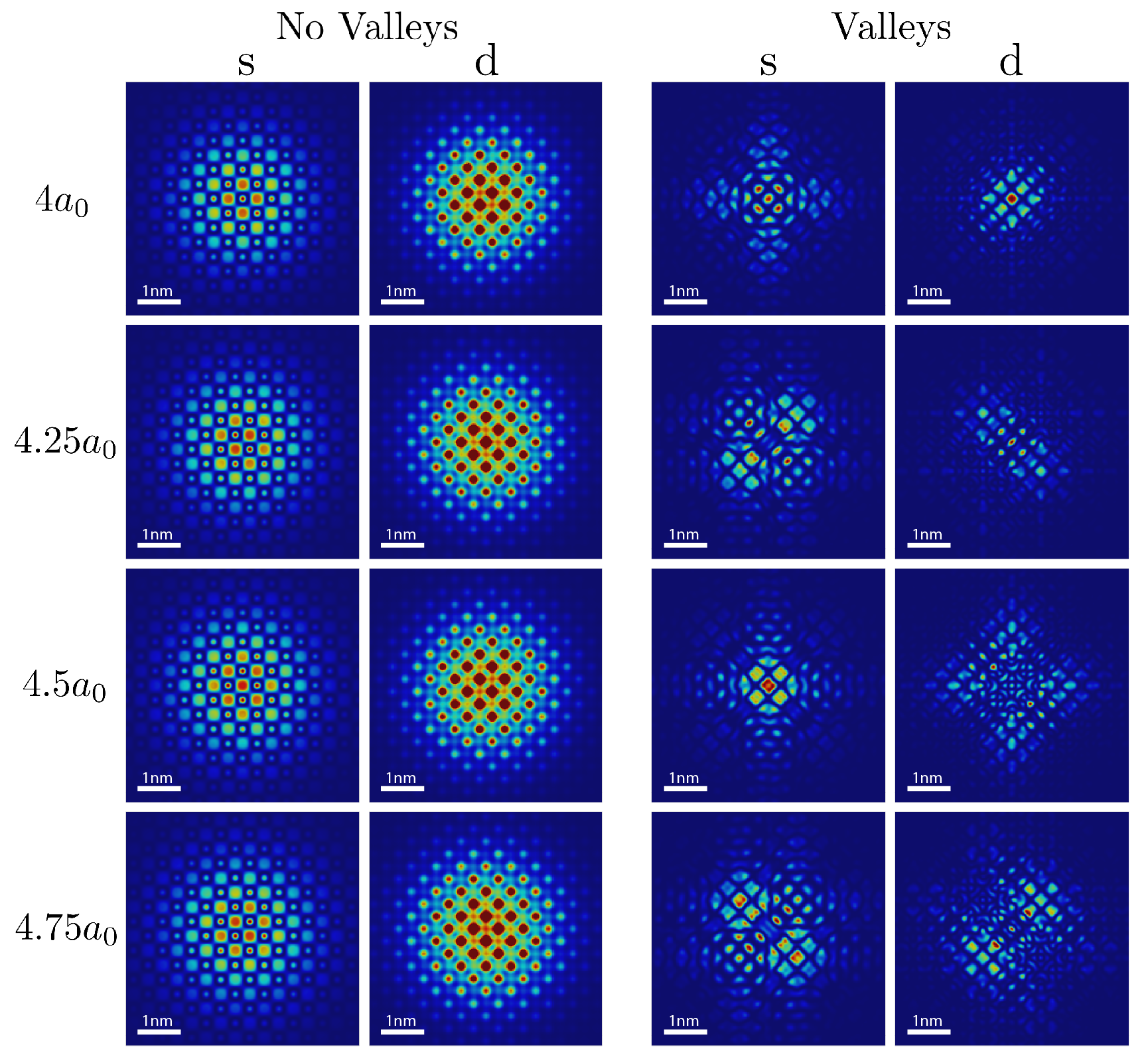}
\caption{The calculated STM images from the charge densities of wave functions from both the valley and no valley KL wave functions are plotted. In the case of valleys, the $d$ type images are drastically different from the $s$ type, while without valleys the features are qualitatively similar in the presence of blurring and saturation. The $d$ type valley  images also capture the experimentally observed alternating of symmetry lines between [110] and [1$\overline{1}$0], with the images switching from one central feature to two off centre features every other atomic plane.}
\label{fig:4}
\end{center}
\end{figure*}

With this analytic form of the wave function, we can simulate the STM images of a P donor in Si not only in the Tersoff-Hamann approximation \cite{Tersoff_PRB_1985}, where the images are simulated by taking the square of the wave function at the apex of a fictitious STM tip as it sweeps across the surface, given by $\abs{\Psi_D(x,y,z_0)}^2$ where $z_0$ is the depth of the P donor under consideration, but also for an arbitrary decomposition of the tip orbital into spherical harmonics, by applying  the appropriate differential operator as specified by Chen's derivative rule \cite{Chen_PRB_1990}. Although the KL model does not include a central-cell correction\cite{Usman_JPCM_2015} or the effects of the Si 2$\times$1 surface reconstruction \cite{Craig_SS_1990}, it manages to reproduce many of the qualitative features of the experimentally observed images. In particular it captures the counter intuitive cyclic sequence of ``butterfly'' and ``caterpillar'' shaped images, with symmetry axes alternating between the [110] and [1$\overline{1}$0] axes as reported in \cite{Usman_NN_2016, Saraiva_PRB_2015} (see supplementary information Figure \ref{fig:S7}). Unlike in \cite{Saraiva_PRB_2015}, where this sequence was found directly from the charge density of the donor wave function (\textit{i.e.}, $s$ type images) evaluated at interstitial planes, here we find the sequence by calculating $d_{z^2-r^2/3}$ type images at atomic planes (Figure \ref{fig:4}).\\ 

Contrary to complex tight-binding simulations (shown in Figure \ref{fig:2}), in this simple analytical form, it is easy to isolate and (artificially) remove the effect of the valleys by setting $v = 0$. The simulated images are plotted in Figure \ref{fig:4} both with and without the valleys, and for  $s$ and $d_{z^2-r^2/3}$ type orbitals. In the no valley case, there is little qualitative change in the images as the depth changes, and the $d_{z^2-r^2/3}$ images are essentially more focused versions of the $s$ images, and become qualitatively similar after the application of blurring. The direct relationship between the features of the $s$ and $d_{z^2-r^2/3}$ images be seen clearly by overlaying a contour plot of the $d_{z^2-r^2/3}$ type images on the $s$ type images (see supplementary information Figure \ref{fig:S6}). \\

In the valley case, however, both $s$ and $d_{z^2-r^2/3}$ images change dramatically as a function of depth, and at a given depth substantially differ in structure. This behaviour agrees with that found via the tight binding simulations in Figure \ref{fig:2}, albeit only qualitatively due to the simplified analytical theory. We can trace the complicated variation of the $d_{z^2-r^2/3}$ images as a function of depth to the valley terms; while the $F_\mu$ are slowly varying functions on the length scales of interest,  the incommensurability of the valley wave vectors with the lattice causes the terms in the sum \[\abs{\Psi_D(\boldsymbol{r})}^2 = \sum_{\mu,\nu,\boldsymbol{G},\boldsymbol{G'}}A_{\mu,\boldsymbol{G}}A_{\nu,\boldsymbol{G'}}^* F_{\mu}(\boldsymbol{r})F_{\nu}^*(\boldsymbol{r}) e^{i \boldsymbol{r}\vdot \left( \boldsymbol{k}_{\mu}+\boldsymbol{G}-\boldsymbol{k}_{\nu}-\boldsymbol{G'} \right) }\] to have completely different phase factors as one moves from one atomic plane to the next, causing different terms to interfere constructively or destructively from plane to plane. As in \cite{Saraiva_PRB_2015} this can also be seen by systematically dropping terms from the KL expression.\\

In conclusion, understanding of the STM images of electron wave functions bounded to points defects in semiconductors is an important component of materials science and engineering at the atomic scale. In this work, we have shown that the observed sensitivity of STM images of P dopants in Si to the quantum mechanical state of the STM tip itself can be attributed to the six conduction band valleys of the indirect Si band structure, by contrasting the tip dependence of STM images simulated in direct and indirect band gap materials. This was further investigated by means of the Kohn-Luttinger model of the donor physics, within which we can artificially remove the effect of the valleys and examine STM images simulated in their absence. The presented results in the context of silicon material will be relevant for the STM measurements on a range of other indirect materials such as Ge, SiGe, and AlSb. Our work provides important new insights for the theoretical understanding of future STM studies where the momentum space features of the sample wave function dictate the measured features.

\noindent
\\
The authors acknowledge useful discussions with Lloyd Hollenberg. This work was supported by the Australian Research Council (ARC) funded Center for Quantum Computation and Communication Technology (CE170100012), and partially funded by the USA Army Research Office (W911NF-08-1-0527). Computational resources were provided by the National Computing Infrastructure (NCI) and Pawsey Supercomputing Center through National Computational Merit Allocation Scheme (NCMAS). This research was undertaken using the LIEF HPC-GPGPU Facility hosted at the University of Melbourne. This Facility was established with the assistance of LIEF Grant LE170100200.

\noindent
\\
The authors declare no competing financial or non-financial interests.
\noindent
\\ \\
The data that supports the findings of this study are available within the article. Further information can be provided upon reasonable request to the corresponding author.

\def\bibsection{\subsection*{\refname}}

\bibliographystyle{naturemag}

\clearpage
\newpage

\onecolumngrid

\noindent
\Large{\textbf{Supplementary Information Document for}}
\noindent
\center {\Large{\textbf{``Influence of sample momentum space features on scanning tunnelling microscope measurements"} }} 

\normalsize

\renewcommand{\thefigure}{\textbf{S\arabic{figure}}}
\renewcommand{\figurename}{\textbf{Supplementary Fig.}}

\setcounter{figure}{0}

\noindent
\begin{figure*}[!h]
\begin{center}
\includegraphics[width=17cm]{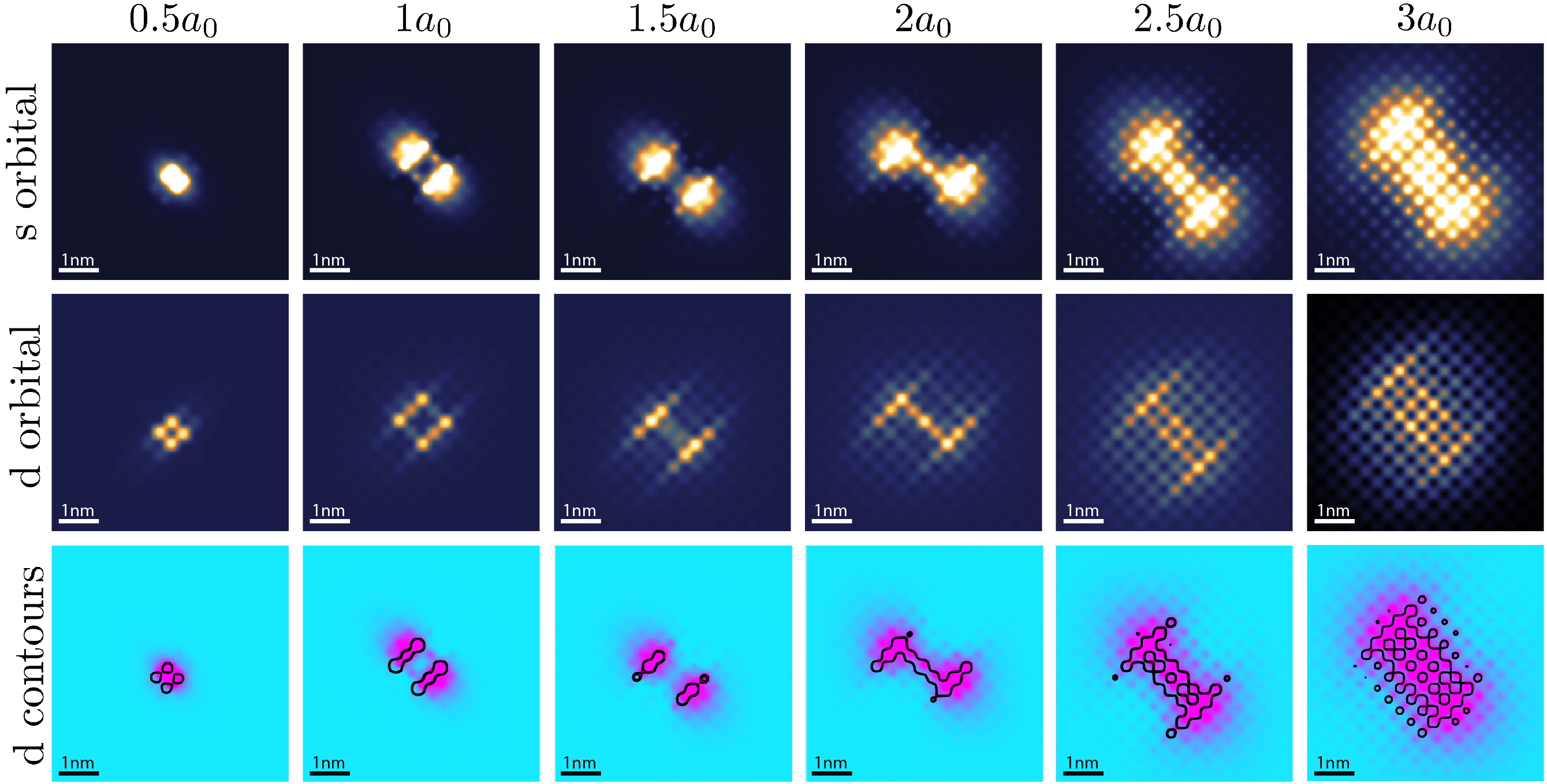}
\caption{STM images of N impurities in GaAs for each of the first 6 planes, both $s$ (top row) and $d$ (middle row) images. Gaussian blurring and colour saturation has been applied to the images which is commensurate with the experimental observations. The contours of the $ d $ images overlaid on top of the $ s $ images (bottom row) which show a good agreement between the two sets of images. }
\label{fig:S1}
\end{center}
\end{figure*}

\begin{figure*}[!h]
\begin{center}
\includegraphics[width=17cm]{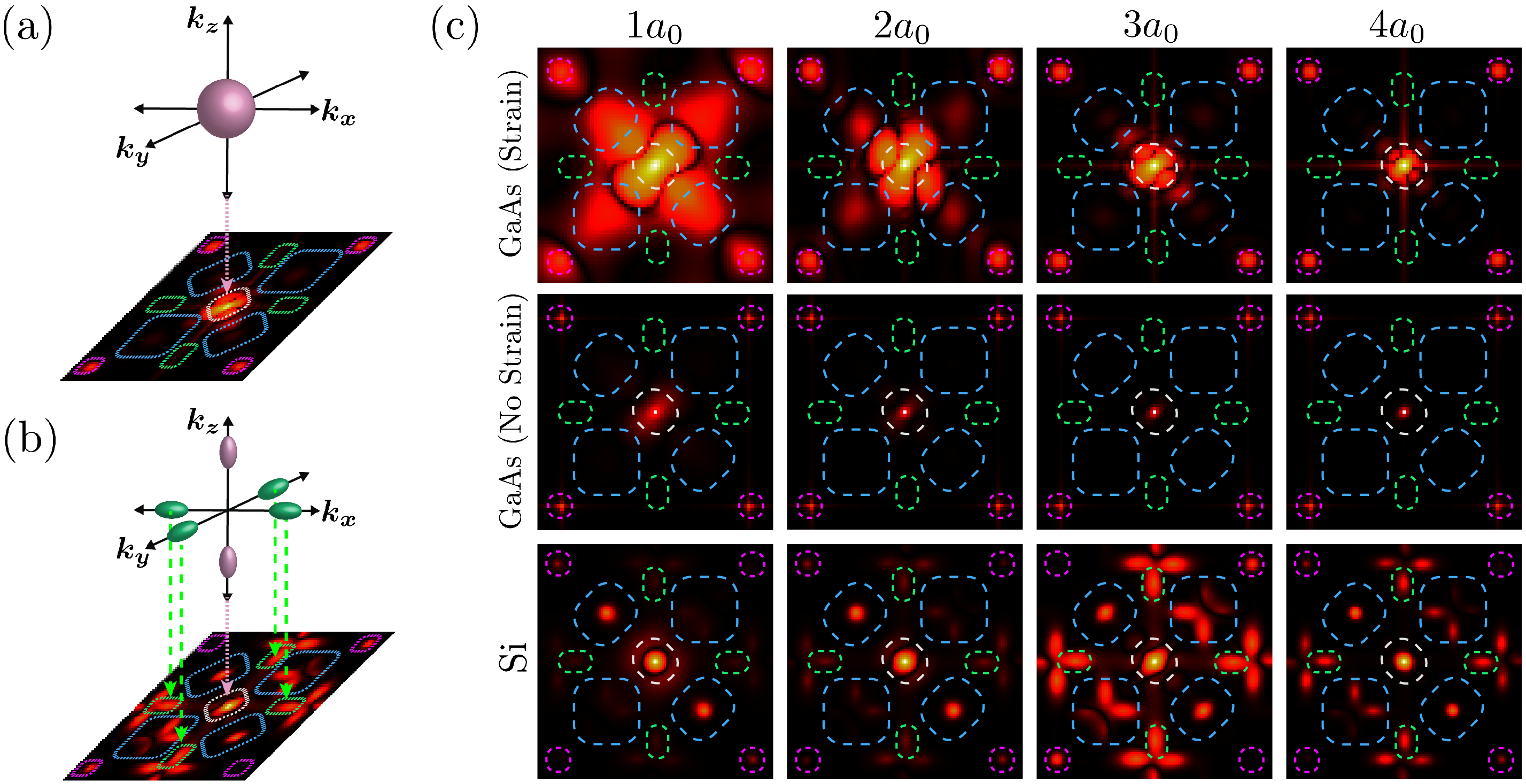}
\caption{(a) Schematic plot indicating the projection of 3-dimensional valley space on 2-dimensional STM image Fourier spectra for a direct bandgap material such as GaAs:N studied in this work. (b) Schematic plot indicating the projection of 3-dimensional valley space on 2-dimensional STM image Fourier spectra for an indirect bandgap material such as Si:P studied in this work. (c) Fourier transforms of STM images of impurity wave functions. Beyond a few layers from the surface the N impurity states have simpler Fourier spectra, consisting only of low frequency components and periodic components at the reciprocal lattice vectors.  The features which occur in the highlighted blue regions in the GaAs:N Fourier spectra are attributed to the lattice strain around the N impurity atom, as they disappear when the strain is artificially turned off and N impurity is placed in unperturbed GaAs lattice. The Fourier spectra of the P donors are more involved. The highlighted features in the Si:P spectra reflect: the low frequency probability envelope and $z$ valleys projection (white), the $x$ and $y$ valleys (green), 2$\times$ 1 surface reconstruction induced features (blue), and the periodic components (pink). As expected from the valley re-population effect, shallow Si:P systems exhibit almost no features in the green regions indicating that $x$ and $y$ valleys are now de-populated. }
\label{fig:S2}
\end{center}
\end{figure*}

\begin{figure*}[!h]
\begin{center}
\includegraphics[width=15.5cm]{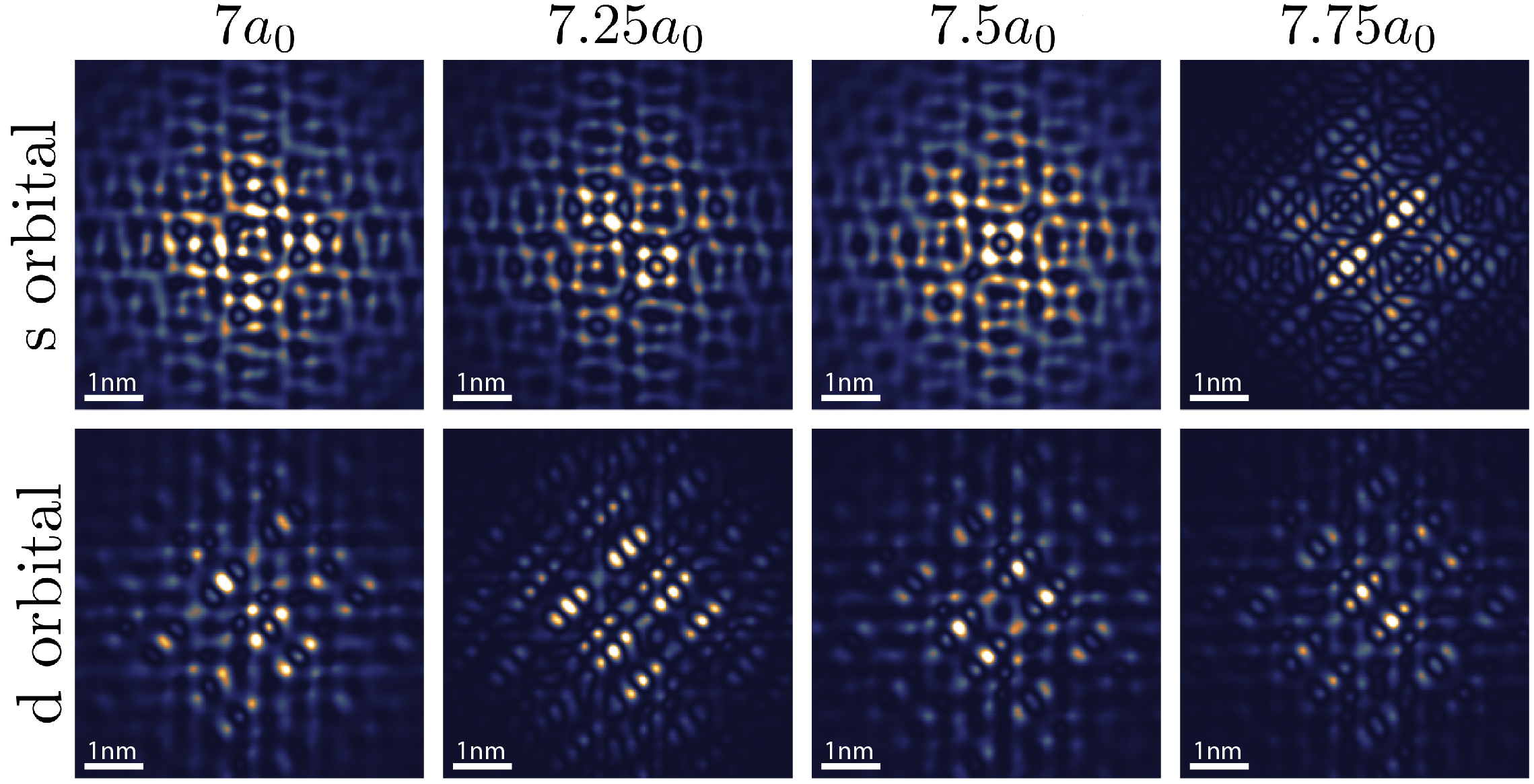}
\caption{The computed STM images of deeper P impurities in Si are shown for both $x$ and $d$ tip orbitals. At these deeper depths the emergence of the full valley structure results in complicated image feature maps, which makes $s$ and $d$ orbital images distinctly different.}
\label{fig:S3}
\end{center}
\end{figure*}

\begin{figure*}[!h]
\begin{center}
\includegraphics[width=15cm]{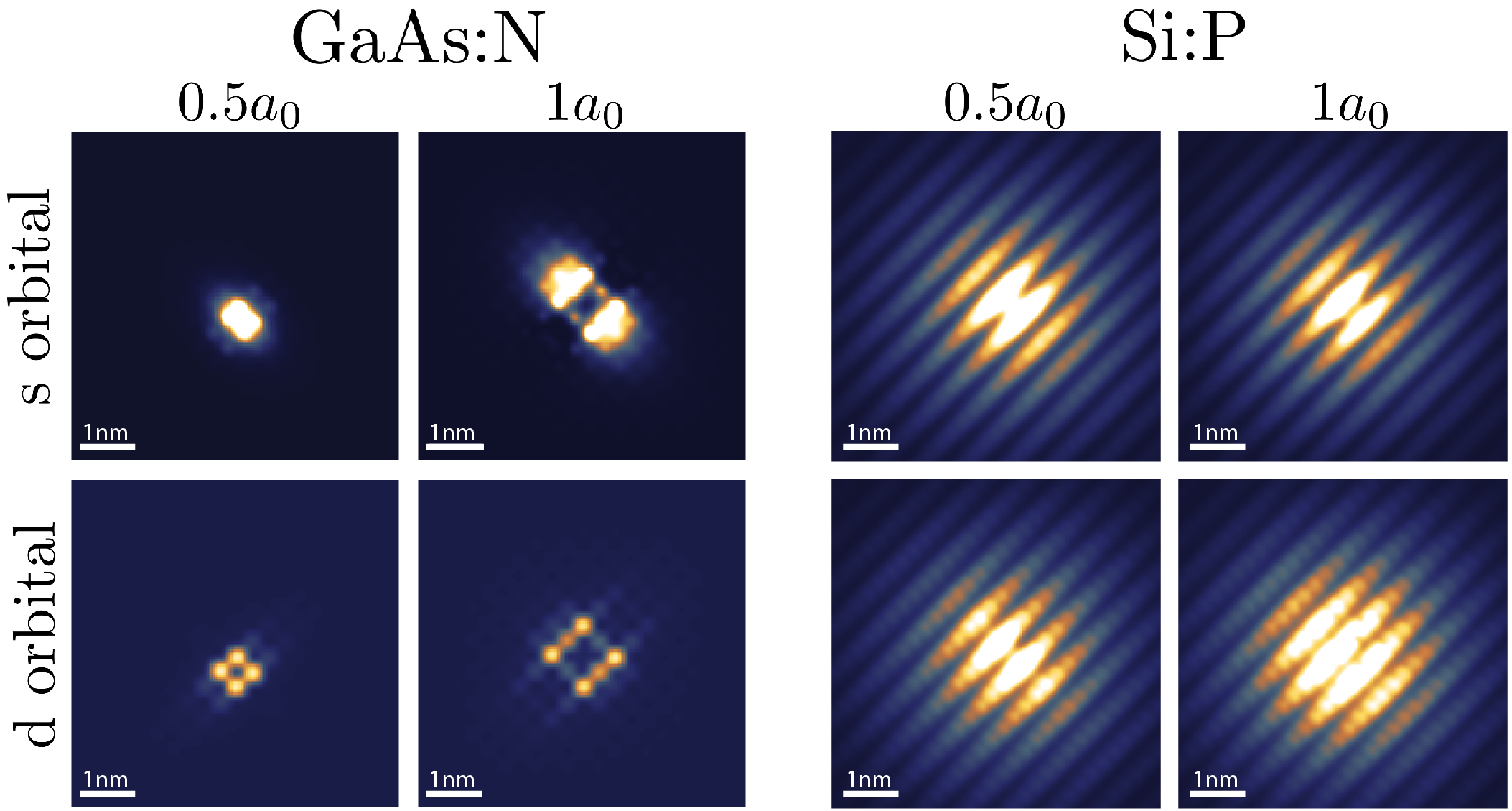}
\caption{The computed STM images of GaAs:N and Si:P systems are plotted for a few shallow depths of N and P impurities. In both cases, the $s$ and $d$ images are qualitatively similar. For Si:P case, the similarity of $s$ and $d$ orbital images is due to significant valley re-population effect at shallow donor depths which leads to donor wave functions comprised of dominantly $z$ valleys. The absence of $x$ and $y$ valleys lead to disappearance of complex image feature maps typically observed at deeper depths.  }
\label{fig:S4}
\end{center}
\end{figure*}

\begin{figure*}[!h]
\begin{center}
\includegraphics[width=15cm]{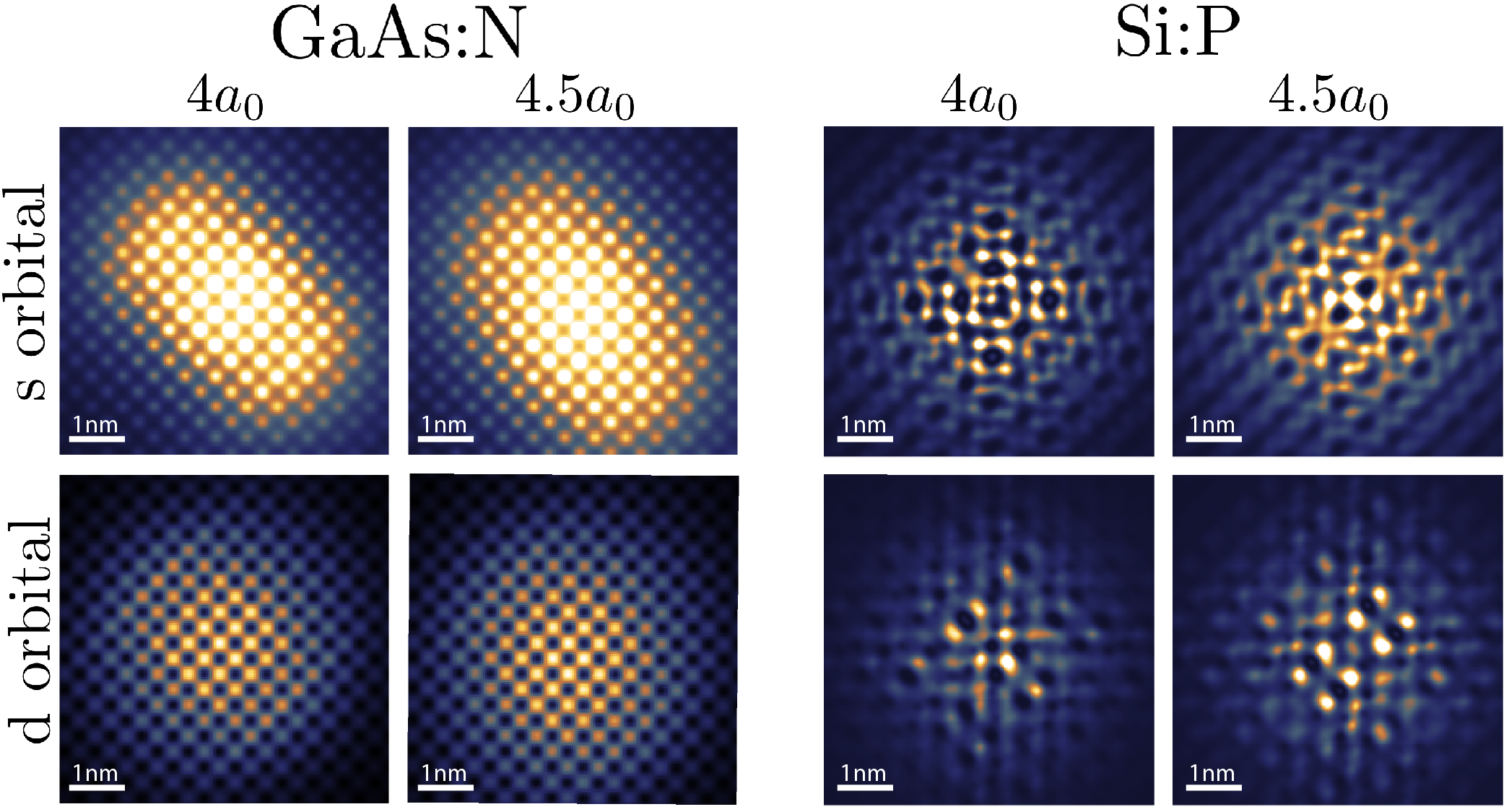}
\caption{The computed STM images of GaAs:N and Si:P for a few selected deep depths. While in the case of GaAs:N the $s$ and $d$ images are qualitatively similar, for Si:P drastic differences is observed between $s$ and $d$ cases.   }
\label{fig:S5}
\end{center}
\end{figure*}

\begin{figure*}[!h]
\begin{center}
\includegraphics[width=16cm]{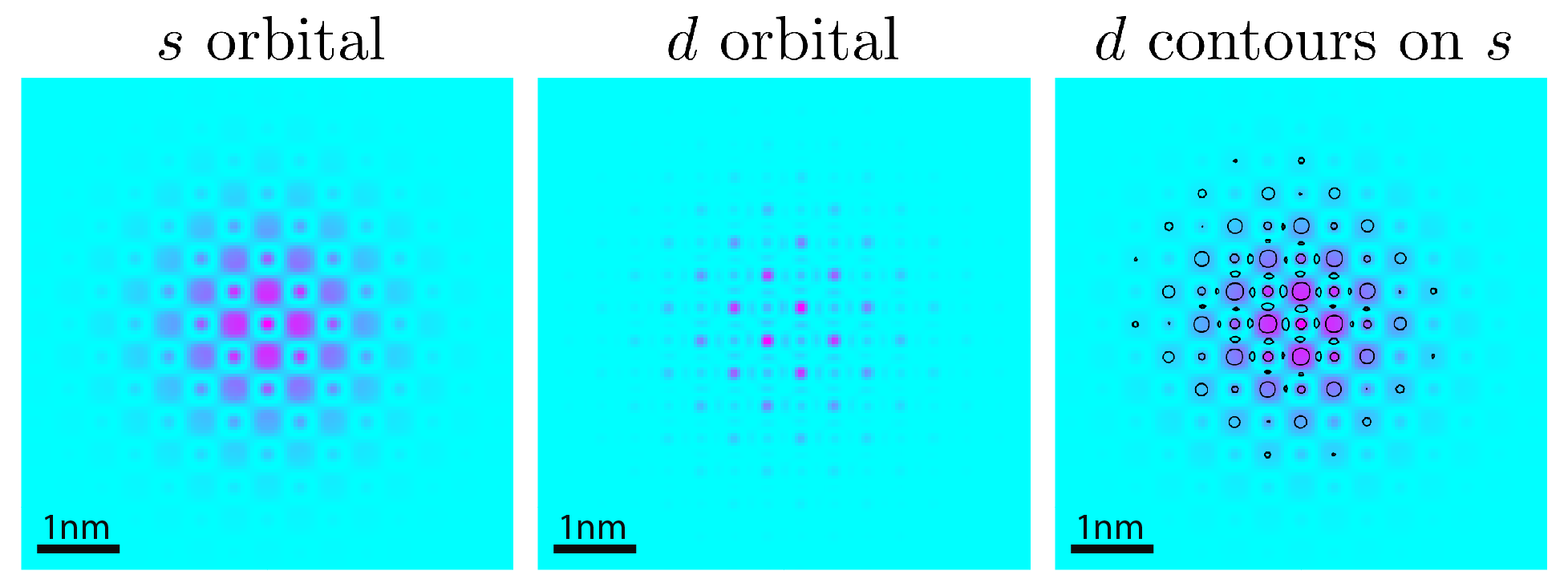}
\caption{The $s$ and $d$ images resulting from a typical Kohn-Luttinger based simulation with the valleys removed. By overlaying the contours of the $d$ image on top of the corresponding $s$ image, we find that almost all features are in direct correspondence, with the main difference being a broadening of the features of the $s$ image. As experimental measurements typically exhibit broadened features due to inherent blurring caused by noise, such qualitative differences will be hard to distinguish in measured images. }
\label{fig:S6}
\end{center}
\end{figure*}

\begin{figure*}[!h]
\begin{center}
\includegraphics[width=16cm]{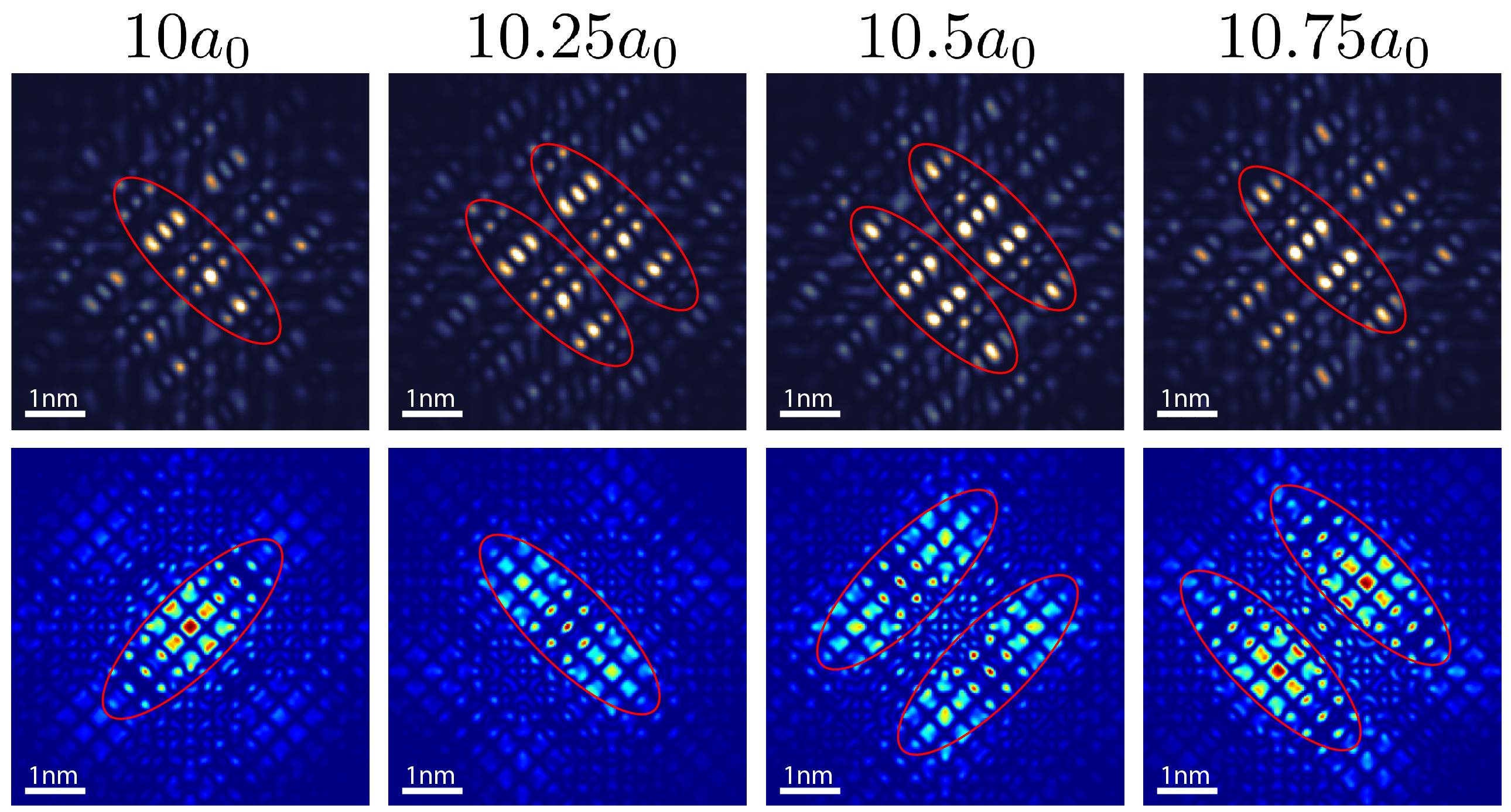}
\caption{The tight-binding and corresponding Kohn-Luttinger (KL) images for a few selected deep donor depths. The images show qualitatively the repeating ``butterfly and caterpillar'' structures in both cases. The KL images display alternating symmetry axes, as earlier reported in \cite{Saraiva_PRB_2015}. }
\label{fig:S7}
\end{center}
\end{figure*}

\end{document}